\documentclass[twocolumn,showpacs,amsmath,amssymb]{revtex4}
\usepackage{graphicx}
\usepackage{dcolumn}
\usepackage{bm}

\begin{document}
\preprint{APS}
\title{Magnetometer suitable for Earth field measurement based on transient atomic response}
\author{L. Lenci}
\author{S. Barreiro}
\author{P. Valente}
\author{H. Failache}
\email{heraclio@fing.edu.uy}
\author{A. Lezama}
\affiliation{Instituto de F\'{\i}sica, Facultad de Ingenier\'{\i}a,
Universidad de la Rep\'{u}blica,\\ J. Herrera y Reissig 565, 11300
Montevideo, Uruguay}

\date{\today}

\begin{abstract}
We describe the development of a simple atomic magnetometer using $^{87}$Rb vapor suitable for Earth magnetic field monitoring. The magnetometer is based on time-domain determination of the transient precession frequency of the atomic alignment around the measured field. A sensitivity of 1.5 nT/$\sqrt{Hz}$ is demonstrated on the measurement of the Earth magnetic field in the laboratory. We discuss the different parameters determining the magnetometer precision and accuracy and predict a sensitivity of 30 pT/$\sqrt{Hz}$.
\end{abstract}


\maketitle
\section{\label{Introduction}Introduction}

Atomic magnetometers are based on the measurement of the Larmor precession frequency of the atomic spin in the magnetic field to be measured. They were first introduced in the sixties \cite{Bloom:1962} before the development of lasers. Present days atomic magnetometers \cite{Kominis:2003,Budker:2007} can rival superconducting quantum interference devices (SQUID), which have been the most sensitive magnetometers for the last three decades (see \cite{Weinstock:1996} for a review of SQUID magnetometers). Very high magnetic field sensitivities have been demonstrated for magnetometers operating at or near the zero of the magnetic field. At larger fields, the performance of the magnetometer is degraded mainly as a consequence of spin-exchange atomic collisions \cite{Happer:1973}. Few atomic magnetometers have been suitably designed for the measurement of the Earth magnetic field and/or geophysical-range fields with high sensitivity. Belfi and coworkers presented the measurement of the variations of the earth magnetic field during some hours \cite{Belfi:2007}. Their device was based on coherent population trapping \cite{Alzetta:1976} with frequency modulation of the laser light. The reported sensitivity of the magnetometer was $45\; pT/\sqrt{Hz}$. Acosta et al. \cite{Acosta:2006} described a technique able to measure geophysical-scale fields with projected sensitivity of $60\; fT/\sqrt{Hz}$ although measurement of the earth magnetic field could not be demonstrated with sensitivity better than $100\; pT/\sqrt{Hz}$, due to environmental magnetic noise. Seltzer and Romalis \cite{Seltzer:2004} have adapted near zero-field spin-exchange techniques to the measurement of an ambient magnetic field by using compensating coils. A sensitivity near $1\; pT/\sqrt{Hz}$ was demonstrated.

In this paper we describe a simple atomic magnetometer setup that, unlike previous work, uses time-domain signal analysis for the measurement of the Larmor frequency of ground state Rb atoms evolving in the Earth magnetic field. In section \ref{Theory} we briefly illustrate the magnetometer principles. The experimental setup is described in section \ref{Experiment}. Section \ref{Measurements} presents the results of test measurements on an artificially generated field and a measurement of the Earth magnetic field.  A critical discussion of the performance of the magnetometer is presented in section \ref{Discussion} followed by the concluding remarks in section \ref{Conclusion}.

\section{\label{Theory} Magnetometer principle}

\begin{figure}[htb]
\includegraphics[width=8cm]{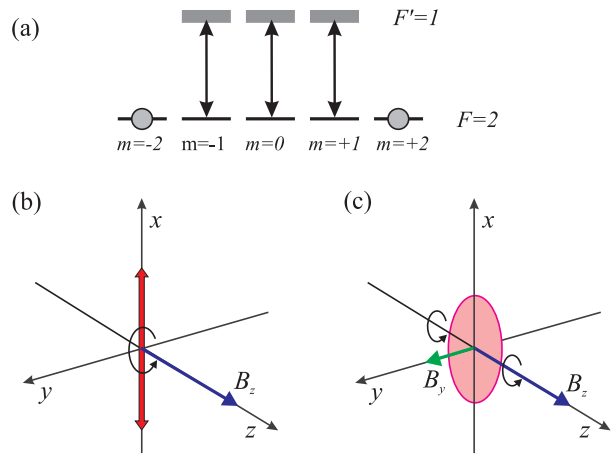}
\caption{\label{fig:preces} (Color online) a) Level scheme for a $F=2$ to $F'=1$ transition with quantization axis parallel to the light polarization. b) Precession of an initially $x$ oriented
atomic alignment around the ambient magnetic field $B_z$
c) Effect of a residual magnetic field $B_y$ present during the alignment preparation; the alignment is spread into a disk in the $xz$ plane that rotates around its $z$ diameter in the presence of the ambient field $B_z$.}
\end{figure}

If an ensemble of atoms, prepared in an anisotropic state of the ground level, is suddenly placed in a magnetic field it will show a transient evolution with oscillations at frequencies which are multiples of the Larmor frequency $\omega_L$ of the atomic ground state. The Larmor frequency relates to the ambient magnetic field $B$ through the relation:

\begin{equation}
\label{eq.Larmor} \omega_L=\mu_B g_F B
\end{equation}
where $\mu_B$ is the Bohr magneton and $g_F$ the ground state hyperfine level Land\'{e} factor which are known with high accuracy \cite{Steck:2010}.  The duration of the transient is determined by the coherence lifetime of the ground state. This transient oscillation provides the basis for the measurement of the Larmor frequency and consequently the ambient magnetic field.

A well known means for preparing the ground level of an atomic sample in an anisotropic state is optical pumping \cite{Happer:1972}. Optical pumping with linearly polarized light usually results in atomic \emph{alignment}. A detailed consideration of atomic alignment in the presence of a magnetic field can be found in \cite{Rochester:2001}. As an example, consider the situation depicted in Fig. \ref{fig:preces}a where a linearly polarized laser field is resonant with a $F=2 \rightarrow F'=1$ atomic dipolar transition. In this figure, the quantization axis has been chosen along the direction of the optical polarization (say $x$). After many absorption and emission cycles, the atomic population will be evenly pumped in the two $m=\pm 2$ Zeeman states. The alignment of the system is here characterized by a well defined direction which is $x$ in the present case. In the presence of an external magnetic field, the alignment precesses around the field at the Larmor frequency $\omega_L$. Notice that if the magnetic field is perpendicular to the alignment (say in the direction $z$) then the alignment reproduces itself after half a precession cycle (see Fig. \ref{fig:preces}b). As the alignment evolves, the absorption of an $x$ polarized light beam is modulated at frequency $2\omega_L$ if the magnetic field is perpendicular to $x$ and at frequencies $\omega_L$ and $2\omega_L$ in the general case.

Our magnetometer uses a sequence of two consecutive time intervals. In the first, the atomic sample is aligned through optical pumping in the absence of magnetic field. In the second interval, the previously prepared atomic alignment freely evolves in the presence of the magnetic field to be measured. A single linearly polarized laser beam, resonant with the atomic transition, is used for both optical pumping and probing the transient evolution of the atomic ground state.

\begin{figure}[htb]
\includegraphics[width=8cm]{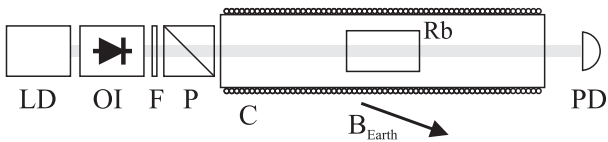}
\caption{\label{fig:setup} Basic scheme of the magnetometer setup. LD: laser diode, OI: optical isolator, F: neutral density
filter, P: polarizer, C: coil and PD: photodetector.}
\end{figure}

The proposed scheme requires the cancellation of the external magnetic field during the optical pumping time. This is achieved with the help of a solenoid oriented roughly parallel to the ambient magnetic field. As it will be discussed below, this compensation does not require the precise a priori knowledge of the ambient magnetic field and can be successfully achieved. After a time of a few milliseconds the system reaches a steady state.  The current in the coil is subsequently switched off during the field measurement time interval.

The functioning of the magnetometer requires the tuning of two parameters. One is the current of the solenoid that cancels the ambient field. Such current is chosen for a purely exponential evolution of the laser transmission during the optical pumping time interval and for maximizing the amplitude of the oscillatory transient during the free evolution time interval. The proper value of the electric current can easily be found by tuning a variable current supply. Nevertheless, this is a critical parameter since a few percent modification in its value results in a significant deterioration of the precession signal.

The second parameter to be carefully chosen is the linear polarization of the laser field. By rotating the polarizer, the field polarization is set to be perpendicular to the ambient magnetic field. Such condition can be met by imposing that the oscillatory transient occurs at frequency $2\omega_L$ with negligible modulation at the subharmonic $\omega_L$. An alternative strategy that we have successfully tested was to let the polarizer in a fixed position and include in the signal processing (see below) the fact that the light transmission evolves at the two frequencies $2\omega_L$ and $\omega_L$. Finally it is worth mentioning that the analysis of the transmission signal for different orientations of the polarizer may lead to a complete knowledge of the vectorial magnetic field.

\section{\label{Experiment} Magnetometer setup}

The experimental setup is sketched in Fig.\ref{fig:setup}. We have used a CW diode laser tuned to resonance with
the $^{87}Rb(F=2 \rightarrow F'=1)$
transition of the D1 line (795 nm). A linear absorption setup with a secondary atomic sample was used
to stabilize the laser frequency on the Doppler absorption
profile. It was tested that a laser frequency variation of the order of
$100\;MHz$ had negligible influence on the magnetometer precision making unnecessary the stabilization of the laser
frequency on a narrower atomic reference signal as a
saturated absorption.

The laser beam was expanded and a $8\;mm$ diaphragm used to
select the center of the beam and obtain an intensity homogeneity better than $10\%$. Neutral density
filters were used to have $25\;\mu W$ radiation power at the atomic cell.
The polarization of the beam was determined with a rotatable linear polarizer.

The glass cell with the Rb vapor is $5\;cm$ long and have windows
of $2.5\;cm$ diameter. It containes both Rb isotopes in
natural abundance and $30\;Torr$ of Ne as a buffer gas. The optimal
conditions for the atomic signal were obtained for an atomic density
of $1.6 \times 10^{11} at./cm^3$ corresponding to a cell temperature
of $55\;^oC$ and a laser absorption of $15 \%$. To avoid a stray
magnetic field from an electric heating system, the cell was heated
with hot water flowing through a silicone tube wrapped around the
cell body. The water temperature was not actively stabilized but
good thermal isolation and large thermal inertia allows a
temperature stability of the cell of about $0.1\;C/hr$.

The cell was placed inside  a $5\;cm$ diameter, $30\;cm$ long  solenoid coaxial with the light beam. The current in the coil originated from a current supply controlled with a signal generator producing square pulses. A reverse polarized diode insured that the electric current is extinguished in 2 $\mu s$.

The photodetector is a linear
photodiode with $1\;MHz$ bandwidth.

\begin{figure}[htb]
\includegraphics[width=8cm]{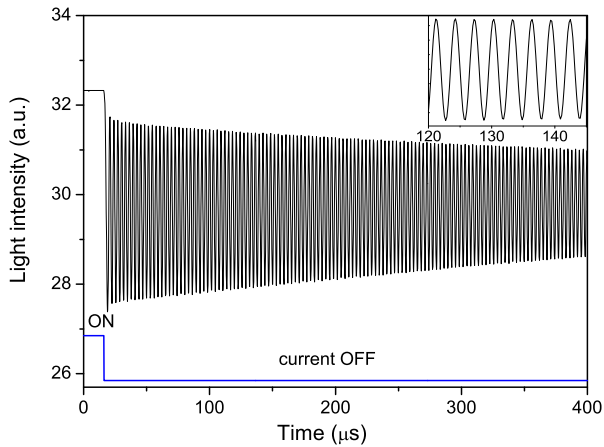}
\caption{\label{fig:trans} Damped oscillation of the transmitted light signal in the presence of the Earth magnetic field used for the determination of frequency $2\omega _L$. Also shown is the time sequence of the current through the coil during the Earth field measurement.}
\end{figure}

\section{\label{Measurements} Measurements}

Fig. \ref{fig:trans} shows a typical transient damped oscillation of the transmitted laser light due to the precession of the atomic alignment in the presence of the magnetic field to be measured.

Different analog or numerical signal processing methods
can be used to extract oscillation frequency from the observed data. We chose to determine the value of $2\omega_L$ by numerically fitting the damped oscillation to the
function:

\begin{equation}
\label{eq.model} A(t)=A_o e^{-\gamma_1 t}cos(2\omega_L t + \phi) + B_o
e^{-\gamma_2 t} + C_o
\end{equation}

Here the damping constant $\gamma_1$ is the decoherence rate of the
sub-levels of the Zeeman manifold, $A_o$ is the absorption
oscillation amplitude, $B_o$ is the amplitude of a damped oscillation background
characterized by the damping constant $\gamma_2$ and $\phi$ a phase
parameter. The use of Eq. \ref{eq.model} has been adopted phenomenologically. It can also be justified from first principles by considering the dynamics of an $F=2$ ground state Zeeman manifold interacting with a resonant linearly polarized optical field in the presence of a constant magnetic field \cite{Valente:2002}. In the determination of the magnetic field from Eq. \ref{eq.Larmor} we have used $\mu_B g_F=2\pi\; 0.699583$ MHz/G \cite{Steck:2010}.

\subsection{\label{TestMeasurements} Test measurements}
Test measurements were initially carried under controlled
conditions on an artificially created magnetic field inside a magnetic shield. A three layers $\mu$-metal
magnetic shield was used to reduce the stray magnetic field produced by many
sources in the laboratory. The residual stray magnetic field inside the
most internal shield was less than $100\; pT$. For the test measurements,
the coil is operated differently than for ambient field measurements. During the atomic
alignment (optical pumping) interval the current through the coil is zero. During the
measurement interval the coil is used to produce a controlled magnetic
field to be measured. This magnetic field, oriented in the direction of the laser propagation, has an estimated inhomogeneity of less than 50 nT along the cell.

\begin{figure}[htb]
\includegraphics[width=8cm]{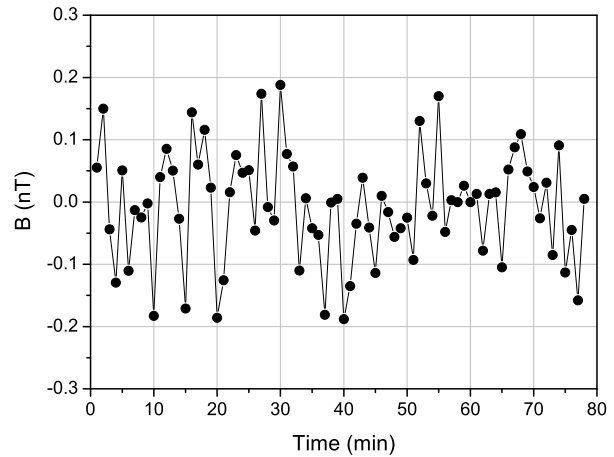}
\caption{\label{fig:noise} Typical dispersion of magnetic field measurements inside a magnetic shield.}
\end{figure}

The transient laser beam transmission through the cell under test conditions with
a magnetic field approximately equal to that of the Earth is similar to that shown in Fig. \ref{fig:trans}. We have checked that the detected signal decay is not affected by laser intensity-broadening. The decay time is mainly determined by the magnetic field inhomogeneity in the sample.

The transient laser absorption signal was acquired with a digital
oscilloscope (Tektronix TDS2232) after averaging $256$ traces with 2048 points each. The
measurement time, a few seconds in our case, was imposed by the slow standard digital
communication protocol between the oscilloscope and the computer. The acquisition time of one transient is however only limited by the atomic coherence decay time ($\sim ms$). A dedicated electronics can, in principle, achieve this limit acquisition rate.

Fig. \ref{fig:noise} shows the variation of the measurements of the magnetic field produced by the coil inside the magnetic shield. The
noise amplitude, typically 400 pT, is mainly determined by the
coil current relative stability estimated to be of the order of $10^{-5}$.

\begin{figure}[htb]
\includegraphics[width=8cm]{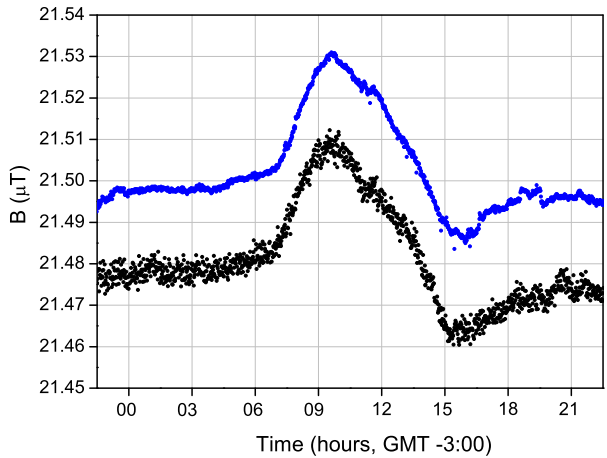}
\caption{\label{fig:earth} (Color online) Earth magnetic field measurement over 24
hours [8/27/2011; 34.919$^{\circ}$S, 56.167$^{\circ}$W] with a proton magnetometer (upper trace) and with the atomic
magnetometer.}
\end{figure}

\subsection{\label{EarthMeasurements} Earth field measurements}

Earth magnetic field measurements were achieved by placing the atomic sample and the surrounding coil in a
separate room free from many sources of magnetic field present in the
laboratory. The laser light was guided from the laboratory to the atomic sample with a 20 $m$ long mono-mode optical fiber (not preserving
polarization). In spite of this precautions, significant magnetic field fluctuations and inhomogeneities were still present. Some of the magnetic field fluctuations are synchronous to the electric power line. Their influence can be reduced by triggering the data acquisition in phase with the power line. However, other uncontrolled
magnetic noise sources due to human activity in the building
persist. In addition, light polarization fluctuations produced in the optical fiber result in amplitude noise in the atomic signal.

Fig. \ref{fig:trans} shows a typical transient of the light transmitted through the atomic cell in the presence of the Earth magnetic field. Earth magnetic field measurements during a $24 hr$ period are presented in Fig. \ref{fig:earth}. Simultaneous measurements
performed with a proton magnetometer (Geometrics G-856AX) are also shown. These
measurements were made few $km$ away from the laboratory and far from
any building since the proton magnetometer was unable to operate under
the typical magnetic field inhomogeneity present inside buildings. We notice that the atomic magnetometer could operate in spite of the large field inhomogeneities probably because its probing volume is more than two order of magnitude smaller than that of the proton magnetometer.

A good correlation between the two measurements is observed in Fig. \ref{fig:earth} showing the daily variation of the Earth field. An absolute
difference of $1\;\mu T$ was present between these two curves that
is attributed to the variation in the magnetic fields
between the two distant measurement points. The result of the
protonic magnetometer  measurement in Fig. \ref{fig:earth} was vertically shifted
for better presentation.

\section{\label{Discussion} Discussion}

We initially address the question of whether the functioning of the atomic magnetometer requires the precise a priori knowledge of the ambient magnetic field for its compensation during the optical pumping interval. We analyze the effect of improper compensation assuming an ambient field oriented along $z$ and an optical polarization along $x$. From Fig. \ref{fig:preces}b we see that imperfect cancelation of the $z$ component of the magnetic field will result in the spreading of the alignment within a disk in the $xy$ plane resulting in a vanishing amplitude of the free evolution transient. For this to occur, it is sufficient that the residual field along $z$ has a Larmor frequency comparable or larger than the optical pumping rate (Hanle resonance width). A suitable value of the current in the magnetic coil can always be found to precisely cancel the $z$ component of the field. However, since the orientation of the magnetic coil will generally not coincide with that of the unknown ambient field, a residual transverse field (in the $xy$ plane) is present during the pumping time interval. The effect of this field is illustrated in Fig. \ref{fig:preces}c where a small residual field component along $y$ is assumed. Its effect is to spread the alignment within a disk in the $xz$ plane. However, such disk, in the presence of the $z$ oriented ambient field, will evolve by rotating around its $z$ diameter giving rise to a significant light transmission modulation signal suitable for magnetic field measurement. We conclude that in order for the atomic magnetometer to operate properly, only a rough initial knowledge of the direction of the ambient field is required to select the orientation of the coil. This was experimentally confirmed by observing that the orientation of the coil axis could be varied by several degrees with no significant degradation of the magnetometer performance.

The measurements shown in Figs. \ref{fig:noise} and \ref{fig:earth} do not represent the
ultimate magnetometer precision. In the case of the Earth field
measurement, several noise sources inside the building degrade the
magnetometer performance, while in the case of operation inside a
magnetic shield the precision is limited by the stability and homogeneity of the magnetic field
to be measured. Since the total effective acquisition time was 0.1 s, we estimate that the data shown in Figs. \ref{fig:noise} and \ref{fig:earth} correspond to a sensitivity of $130\; pT/\sqrt{Hz}$ and $1.5\; nT/\sqrt{Hz}$ respectively.

The limit magnetometer precision can be estimated from the
uncertainty in the determination of frequency $2\omega_L$
from the atomic signal shown in Fig. \ref{fig:trans}. This uncertainty is taken as the mean square root time deviation $\delta T$ of the experimental data from the best fitting curve.  The signal to noise ratio is $S/N=T/{\delta T}$ where $T$ is the oscillation period. The uncertainty $\delta$ in determination of $\omega_L$ is given by:

\begin{equation}
\label{eq.noise} \frac{\delta }{\omega_L} = \frac{1}{\sqrt{n}.S/N}
\end{equation}
where $n$ is the measured number of oscillation periods. An
estimation of $n$ is given by $n\simeq\frac{\omega_L}{\gamma_1}$ where $\gamma_1$ is the transient damping rate.

A lower limit of $\gamma_1$ in our buffered cell was measured to be
$\gamma_1 \leq 50 Hz$ \cite{Failache:2003}. Using this value together with the experimentally achieved S/N ratio $\simeq 3 \times 10^{3}$, we extrapolate a magnetometer sensitivity of $\delta B = 30\ pT/\sqrt{Hz}$.  Enhancement of this figure is in principle possible through improvements of the signal to noise ratio.

In the results presented in this work, the non-linear Zeeman effect was not taken into account. For our experimental conditions, this effect has a negligible influence on the achieved precision. We note that under conditions approaching the limit resolution of the magnetometer, the consequences of the non-linear Zeeman effect would be observable. This effect can  be accounted for in principle by generalizing the expression in Eq. \ref{eq.model} to allow for the presence of modulation sidebands \cite{Acosta:2006}.

The accuracy of our magnetometer is essentially limited by the effect of atomic collisions. While the collisions with the buffer gas are estimated to play a negligible role, we believe that the performance of the magnetometer is affected by  $Rb-Rb$ spin-exchange
collisions \cite{Balling:1964}. We have observed a temperature dependent long term drift of $\frac{\delta B}
{\delta T} \;\sim\;2\;nT/^oC$. This drift is considered to be the consequence of small changes in the atomic density in response to temperature changes and consequently in the rate of spin-exchange collisions. A similar shift was measured by Balling et al. \cite{Balling:1964}.

\section{\label{Conclusion} Conclusion}

We have built an atomic magnetometer suitable for measurements of geophysical magnetism. The magnetometer uses time domain signal processing to determine the Larmor frequency of ground state Rb atoms in the presence of the ambient field. The very simple experimental setup does not require the precise a priori knowledge of the ambient field. The orientation and tuning of the experimental apparatus is easily achieved and is robust again small changes in the experimental conditions. The sensitivity of our magnetometer is comparable to that reported for other atomic magnetometers
used to measure the Earth magnetic field. The extension of our work towards the development of a vectorial magnetometer is currently considered.\\

\begin{acknowledgments}
We wish to thank L. S\'{a}nchez for providing the commercial proton magnetometer and A. Saez for help with the experiment setup. This work was supported by CSIC, ANII and PEDECIBA (Uruguayan agencies).
\end{acknowledgments}


\end{document}